\documentclass[a4paper,fleqn,usenatbib]{mnras}

\usepackage[T1]{fontenc}
\usepackage{ae,aecompl}

\usepackage{graphicx}   
\usepackage{amsmath}    
\usepackage{amssymb}    
\usepackage{color}

\title{Non-parametric Morphologies of Mergers in the Illustris Simulation}

\author[L. A. Bignone et al.]{
L. A. ~Bignone$^{1}$\thanks{E-mail:lbignone@iafe.uba.ar},
P. B. ~Tissera$^{ 2, 3}$,
E. ~Sillero$^{4}$,
S. E. ~Pedrosa$^{1}$,
L. J. ~Pellizza$^{5}$
\newauthor
and D. ~Garc\'ia ~Lambas$^{4}$
\\
$^{1}$Instituto de Astronom\'ia y F\'isica del Espacio (IAFE, CONICET-UBA), C.C. 67 Suc. 28, C1428ZAA Ciudad de Buenos Aires, Argentina.\\
$^{2}$Departamento de Ciencias F\'isicas, Universidad Andres Bello, Av. Republica 220, Santiago, Chile.\\
$^{3}$Millennium Institute of Astrophysics, Av. Republica 220, Santiago, Chile\\
$^{4}$Instituto de Astronomia Te\'orica y Experimental (CONICET-UNC), Laprida 925, Cordoba, Argentina.\\
$^{5}$Instituto Argentino de Radioastronom\'ia, CONICET, Camino Gral. Belgrano km 40, Berazategui, Prov. de Buenos Aires, Argentina\\
}

\date{Accepted XXX. Received YYY; in original form ZZZ}

\pubyear{2016}

\begin{document}
\label{firstpage}
\pagerange{\pageref{firstpage}--\pageref{lastpage}}
\maketitle

\begin{abstract}

We study non-parametric morphologies of mergers events in a cosmological
context, using the Illustris project. We produce mock g-band images
comparable to observational surveys from the publicly available Illustris
simulation idealized mock images at $z=0$. We then measure non parametric
indicators: asymmetry, Gini, $M_{20}$, clumpiness and concentration for a set
of galaxies with $M_* >10^{10}$ M$_\odot$. We correlate these automatic
statistics with the recent merger history of galaxies and with the presence of
close companions. Our main contribution is to assess in a cosmological
framework, the empirically derived non-parametric demarcation line and
average time-scales used to determine the merger rate observationally. We
found that 98 per cent of galaxies above the demarcation line have a close
companion or have experienced a recent merger event. On average, merger
signatures obtained from the $G-M_{20}$ criteria anticorrelate clearly with
the elapsing time to the last merger event.  We also find that the asymmetry
correlates with galaxy pair separation and relative velocity, exhibiting the
larger enhancements for those systems with pair separations  $d < 50$ h$^{-1}$
kpc and relative velocities $V < 350$ km s$^{-1}$. We find that the $G-M_{20}$
is most sensitive to recent mergers ($\sim0.14$ Gyr) and to ongoing mergers
with stellar mass ratios greater than 0.1. For this indicator, we compute a
merger average observability \protect {time-scale} of $\sim0.2$ Gyr, in
agreement with previous results and demonstrate that the morphologically
derived merger rate recovers the intrinsic total merger rate of the simulation
and the merger rate as a function of stellar mass.

\end{abstract}

\begin{keywords}

cosmology: galaxy formation -- galaxies: interactions -- galaxies: structure -- galaxies: evolution -- methods: numerical

\end{keywords}

\defcitealias{lotz_new_2004}{LPM04}



\section{Introduction}

Galaxy mergers are of fundamental importance in the formation and
evolution of galaxies, especially in the  $\Lambda$ Cold Dark Matter ($\Lambda$CDM)
cosmology where structure grows hierarchically
\citep[e.g.][]{white_core_1978}. Mergers have an important effect in the
mass assembly of galaxies \citep{guo_galaxy_2008,genel_halo_2009}, the star
formation history \citep{mihos_gasdynamics_1996,somerville_nature_2001}, the
establishment of galaxy morphologies, internal structures
\citep{mihos_dense_1994,johnston_fossil_1996,
naab_statistical_2003,bell_accretion_2008} and the growth and accretion of gas
by supermassive black holes \citep{hopkins_unified_2006}. Understanding the
role of mergers in the formation of galaxies and their relative importance in
comparison to other, more continuous processes, such as cold gas and dark
matter accretion is a key challenge for galaxy formation models. 

A first step to study the role of mergers in galaxy evolution is the
estimation of the merger rate by counting the observed number of events. There
are several approaches for the identification of such systems. Galaxy pairs
with close projected angular separations and low line-of-sight relative radial
velocities, for example, can be considered suitable merger candidates
\citep{barton_tidally_2000, patton_dynamically_2002, lambas_galaxy_2003,
lin_deep2_2004,de_propris_millennium_2005}. An alternatively method is the
identification of morphologically disturbed galaxies, either through visual
inspection
\citep{kampczyk_simulating_2007,bundy_mass_2005,brinchmann_hubble_1998} or by
quantitative measurements of non-parametric morphological statistics such as
the Gini coefficient ($G$), the second-order moment of the brightest 20
\protect{per cent} of the light \citep[$M_{20}$,][hereafter
LPM04]{lotz_new_2004} and the  $CAS$ system formed by the combination of
concentration ($C$) \citep{wu_quantitative_1999,bershady_structural_2000,conselice_asymmetry_2000}, rotational asymmetry ($A$)
\citep{abraham_galaxy_1996,conselice_asymmetry_2000}, and clumpiness ($S$) \citep{isserstedt_spate_1986,takamiya_galaxy_1999,conselice_relationship_2003}.

Different methods of merger candidate selection might be sensitive to
different merger stages. For example, selecting close pairs imply a higher
probability of capturing galaxies in the pre and early merger stages, while
morphological disturbances methods are more sensitive to  pre, ongoing and
post merger stages \citep{lotz_major_2011}. It is not surprising then, that
observational constraints of the merger fractions can differ by up to an order
of magnitude and yield very different redshift evolution depending on
the method adopted
\citep[e.g.][]{lotz_evolution_2008,de_ravel_vimos_2009,bundy_greater_2009,
lopez-sanjuan_galaxy_2009,conselice_structures_2009}.

In order to properly obtain merger rates from observed merger fractions, an
observability time-scale (i.e. the average time during which a merging system
would be identified as such by applying certain criteria) has to be adopted.
This time-scale might be sensitive to a wide variety of factors such as the
merger selection criteria, the interacting galaxy  properties (e.g. mass
ratio, gas fraction, orbital parameters) and observational parameters (e.g.
viewing angle, resolution, observed wavelength, S/N ratio).
\citet{lotz_galaxy_2008} used a series of numerical simulations of equal-mass
interacting galaxy pairs to constraint the observability time-scales for a
variety of non-parametric morphological statistics. The simulations were
processed trough a radiative transfer code that resulted in realistic mock
images of the interacting galaxies at different merger stages. Then the images
were  used to study the dependence of morphological statistics on the merger
stage, viewing angle, orbital parameters and gas properties. Similar methods
were used to study the effect of merger mass ratio \citep{lotz_effect_2010-1}
and gas fraction \citep{lotz_effect_2010}.

However, the use of isolated merger simulations to derive the observability
time-scale comes with a significant limitation, they do not account for
the cosmological context of galaxy formation. Instead, the observability
time-scales for each merger parameter set (e.g. mass ratios, gas fractions)
has to be weighted by the probability distribution of such parameters.
Typically, these  distributions  are poorly constraint observationally and
have to be derived from independent, cosmological-scale simulations. Using
this methodology, \citet{lotz_major_2011} successfully reconciled the uneven
observational merger rates at $z < 1.5$, and were able to differentiate the
rates for major and minor mergers. They also compared the evolution of the
merger rates with theoretical predictions of galaxy evolution models, finding
an excellent agreement for the major merger rate. Conversely, the total merger
rate (minor and majors combined) derived from the $G-M_{20}$ diagnostic,
was an order of magnitude higher that the rate predicted by the cosmological
simulations which were used to derive the distributions of merger parameters.
This result suggested  a possible underestimation of the $G-M_{20}$
observability time-scale derived from isolated interacting pair simulations
\citep{lotz_galaxy_2008, lotz_effect_2010-1}.

An appealing alternative is the study of the non-parametric morphology
indicators of mergers directly selected from cosmological hydrodynamical
simulations. In this simulated cosmological context, interacting galaxies
cover a wide range of stellar masses, gas fraction, environments, mass ratios
and orbital parameters that closely resemble what observational studies of
such systems must encounter. Recently \citet{snyder_diverse_2015} used a set
of 22 zoomed-in galaxies to quantify the morphological evolution at $z > 1$,
including the morphological effects of mergers. While zoomed-in simulations
certainly capture the cosmological context of galaxy formation, they still
constitute a small sample size that explores a limited parameter space. A
solution to this limitation is to explore non-parametric morphological
statistics of a galaxy catalogue selected from a large simulated volume. This
clearly represents a significant technical challenge which only recently has
became possible to tackle. At $z=0$, \citet{snyder_galaxy_2015} studied the
Gini-$M_{20}$ morphology of 10808 galaxies from the Illustris simulation
\citep{vogelsberger_introducing_2014-1}. These authors found that the
morphological distribution of simulated galaxies agreed well with
observations, and that important relationships such as the connection between
morphological type and stellar mass (M$_*$) and morphological type and star
formation rate, follow the trends reported by different galaxy surveys.

It is also important to point out that hydrodynamical simulations face
important challenges in reproducing the complicated physical processes
involved in galaxy formation, and that further improvements in the modelling
of certain aspects which regulate star formation such as cooling rates, gas
inflows and outflows and feedback should be expected. In the case of the
Illustris project, \citet{sparre_star_2015} found that the observed
relationship between star formation rate and stellar mass at $z=0$ and 4 are
well reproduced, but not at intermediate redshifts where the normalization of
the relationship is too low. Numerical resolution is also an important factor
to consider. \citet{sparre_star_2015} found a paucity of strong starbursts in
the Illustris simulation which can affect the appearance of mergers where
induced star formation is expected. In fact, this resolution effect was
further studied by \citet{sparre_zooming_2016} where zoom-in simulations of
major mergers at 10-40 times higher mass resolution than Illustris where more
successful at generating starbursts using the same physics model. Numerical
resolution can also impact the mock images generated from cosmological
simulations given that the mass of simulated stellar particles can be 2-3
orders of magnitude higher than actual star forming regions, this can affect
the galaxy appearance \citep{torrey_synthetic_2015} and also their colours and
luminosities \citep{trayford_colours_2015}. These caveats affect all currently
large volume simulations of this kind since it is not yet possible to simulate
such large volumes at higher resolutions.

In this work we take full advantage of the large cosmological volume of the
Illustris simulation to derive a statistically significant number of \protect
{non-parametric} morphological indicators of galaxies subject to diverse
environmental situations, including isolated system, merging and interacting
pairs. From the publicly available mock images of the Illustris simulation
\citep{torrey_synthetic_2015} we select a sample of galaxies with $M_* >
10^{10}$~M$_\odot$ at $z=0$ to study the ability of Gini, $M_{20}$,
concentration, clumpiness and asymmetry to successfully classify close pairs,
minor mergers and major mergers. We analyse the effectiveness of the
empirically derived $G-M_{20}$ and $CAS$ merger diagnostics to distinguish
between normal and interacting galaxies in the simulation. Finally, we attempt
to reconcile the intrinsic merger rate of the simulation with the merger rate
derived using the same techniques often used in observational studies
\citep[e.g][]{lotz_evolution_2008,lotz_major_2011}.

The paper is organized as follows. In section 2 we briefly describe the
Illustris simulation and the galaxy samples selected for  our analysis. In
section 3 we explain the procedure applied to the mock images.  In Section 4
we measure the effectiveness of the merger diagnostics in selecting different
populations of interacting  galaxies or recent merger events. In section 5 we
explore the merger rate of the simulation in the light of non-parametric
morphological studies. Finally, in section 6 we discuss our results a present
our conclusions.

\section{Simulated galaxy samples}

In the following sections we provide a brief description of the Illustris
simulations, the galaxy catalogues and the mock images. We also include a
description of the galaxy subsamples defined for our analysis.

\subsection{Overview of the Illustris Simulation}

The Illustris project
\citep{vogelsberger_introducing_2014-1,genel_introducing_2014} consists of  a set of
large-scale hydrodynamical cosmological simulations with periodic box 106.5
Mpc a side, run with the quasi-Lagrangian \textsc{arepo} code
\citep{springel_e_2010}. The galaxy formation model
\citep{vogelsberger_model_2013} includes gas cooling and photo-ionization,
star formation and ISM models, stellar evolution  (gas recycling and
chemical enrichment), stellar supernova feedback and supermassive black holes
with quasar-mode and radio-mode feedback \citep{sijacki_unified_2007,sijacki_illustris_2015}.

The main simulation of the project, Illustris-1 (hereafter, I-1), initially
has 1820$^{3}$ gas cells and 1820$^{3}$ dark matter (DM) particles. The
initial mass of gas elements is $1.26 \times 10^6$~M$_\odot$, while for DM
particles the mass is $6.26 \times 10^6$~M$_\odot$. The I-1
simulation follows structure and galaxy formation across 136 snapshots,
culminating at $z = 0$ and has been shown to reproduce many of the key
observed trends in the local Universe \citet{vogelsberger_introducing_2014-1},
with some discrepancies related to the stellar ages of low mass ($M_* \la
10^{10.5}$ M$_\odot$) galaxies and the quenching of massive galaxies.

The Illustris project adopted the following set of cosmological parameters:
$\Omega_m = 0.2726$, $\Omega_\Lambda = 0.7274$, $\Omega_b = 0.0456$, $\sigma_8
= 0.809$, $n_s = 0.963$ and $h = 0.704$, which are consistent with the
Wilkinson Microwave Anisotropy Probe \mbox{(WMAP)-9} measurements \citep{hinshaw_nine-year_2013}.

\subsection{Galaxy catalogue and mergers trees}

In the I-1 simulation, DM halos were identified using the standard \mbox
{friends-of-friends} (FoF) algorithm \citep{davis_evolution_1985} with linking
length of 0.2 times the mean particle separation and a minimum number of 32 DM
particles. Baryonic elements were assigned to the FoF group of the closest DM
particle. Gravitationally bound substructures within the FoF groups were
identified using the \textsc{subfind} algorithm
\citep{springel_populating_2001,dolag_substructures_2009} resulting in
$4,366,546$ individual subhalos at $z=0$. Subhalos with M$_{*} \simeq 10^{10}$
M$_\odot$ have approximately $30,000$ gas cells, $40,000$ DM particles, and
$10,000$ star particles. We point out that stellar masses used in this paper
are those   obtained from the \textsc{subfind} catalogue, without considering
truncation at any radius.

From the halo and subhalo group catalogues, \citet{rodriguez-gomez_merger_2015} constructed the corresponding merger trees using the newly
developed \textsc{SubLink} code. They argued that in order to avoid problems
caused by the way halo finders distribute mass between substructures, a robust
estimation of the mass ratio of galaxies in a merger event can be obtained by
taking the two progenitor masses at the moment when the secondary progenitor
reaches its maximum stellar mass.  Throughout the rest of this paper, we refer
to merger events taken from the \textsc{SubLink} merger trees. Their mass
ratios are computed according to the above definition.

\subsection{Mock observations}
\label{sec:mock_observations}

\citet{torrey_synthetic_2015} present a method to generate synthetic images
and integrated spectra for galaxies in the Illustris project. They employed the radiative
transfer code \textsc{sunrise} \citep{jonsson_sunrise:_2006, jonsson_high-resolution_2010}
to assign a full spectral energy distribution (SED) to each
star particle and to generate images of arbitrary field-of-view (FOV)
and pixel size for different camera orientations with respect to a galaxy.
The SEDs were  calculated by assuming the single-age stellar populations models by
\textsc{starburst99}
\citep{leitherer_starburst99:_1999,leitherer_library_2010}. A simplified empirical dust model of
\citet{charlot_simple_2000} was also adopted. 

The Illustris project made available through its online
database\footnote{\url{http://www.illustris-project.org/data/}} 6978 mock
observations for galaxies with $M_* > 10^{10}$ M$_\odot$ at $z=0$. Each galaxy
was imaged with four (256 x 256 pixels) cameras orientated in four different
viewing angles which were randomly aligned with respect to the rotational axis
of the galaxy. Each camera was placed 50 Mpc away from the galaxy centre and
the FOV was set to 10 times the stellar half mass radius for the galaxy. It is
important to point out that the mock observations include not only light from
the chosen subhalo, but also from all other structures belonging to the same
halo that fall within the FOV. Because non-parametric indicators can be
affected by the light from projected close companions, this kind of images are
specially suitable to perform the kind of morphological perturbation studies
we propose here.

These images constitute idealized observations because they do not include
noise, camera point-spread-function (PSF) blurring or contamination from
foreground or background sources. They are meant to be used as a base to
compare with observations from different telescopes. Hence, the images must be
degraded according to the particular characteristics of the observations to be
used to confront them. In this paper, we focus on mock rest-frame g-band
morphologies that can be compared to optical and observed morphologies from
the ground. In section \ref{sec:image_analysis}, we describe how the mock g
observations are degraded in order to approximately match real observations.

Our mock image sample has been cleaned of an artefact produced by  the image
generation algorithm which resulted in some empty images. The issue occurred
whenever the halo centre position and the subhalo position were a periodic
boundary away from one another (private communication). We found and removed
99 such subhalos resulting in a final sample of 6879 galaxies.

\subsection{Galaxy samples} 

From the \textsc{SubLink} merger trees, we define a subsample of major merger
remnants made up by galaxies at $z=0$ that experienced at least one major
merger in the previous 2 Gyr. We consider a merger as major when the stellar
mass ratio ($\mu_*$) between the secondary and principal components is larger
than 0.25. We also define a subsample of minor merger remnants composed by
galaxies at $z=0$ that experienced at least one minor merger in the last 2 Gyr
and no major merger. We consider a merger as minor when $\mu_* < 0.25$.
Subhalos included in the merger trees have at least 20 resolution elements
between gas and stars \citep{rodriguez-gomez_merger_2015} which gives a
minimum mass of about $2.6 \times 10^7$ M$_\odot$ and a minimum $\mu_* \sim
0.001$ for $10^{10}$ M$_\odot$ descendants.. The total major and minor
merger remnants subsamples have 322 and 3784 individual galaxies,
respectively.

We also define a subsample of 753 close galaxy pairs. We limit the sample to
those galaxies having a companion with a stellar mass greater than 10$^8$
$h^{-1}$ M$_\odot$ at a distance $d \leq 20$ h$^{-1}$ kpc. At these close distances,
galaxies are more likely to constitute an \protect{on-going} merger or to
experience disturbed morphologies due to gravitational interaction
\citep[e.g.][]{lambas_galaxy_2003, perez_galaxy_2006}. However, not all of
these pairs actually constitute mergers. In order to better constrain the
\protect{on-going} merger sample, we also consider the relative velocities
between the members of the pairs.

Finally, we define a subsample of 5090 distant pairs formed by galaxies having
a companion with a stellar mass greater than 10$^8$ $h^{-1}$ M$_\odot$
within the range $20 < d \leq 100$ h$^{-1}$ kpc. We point out that even if
they do not constitute merger events, galaxies in the distant pair subsample
might show disturbed morphologies \citep{ellison_galaxy_2008,ellison_galaxy_2013}.

There are an additional 582 galaxies that do not fulfil any of the criteria
defined above. They have not experienced any merger in the last 2 Gyr nor do
they have any companion closer than 100 $h^{-1}$ kpc . This subsample
constitute a useful control sample of \textit{unperturbed} galaxies.

\begin{figure*}
    \centering
    \includegraphics[width=0.7\textwidth]{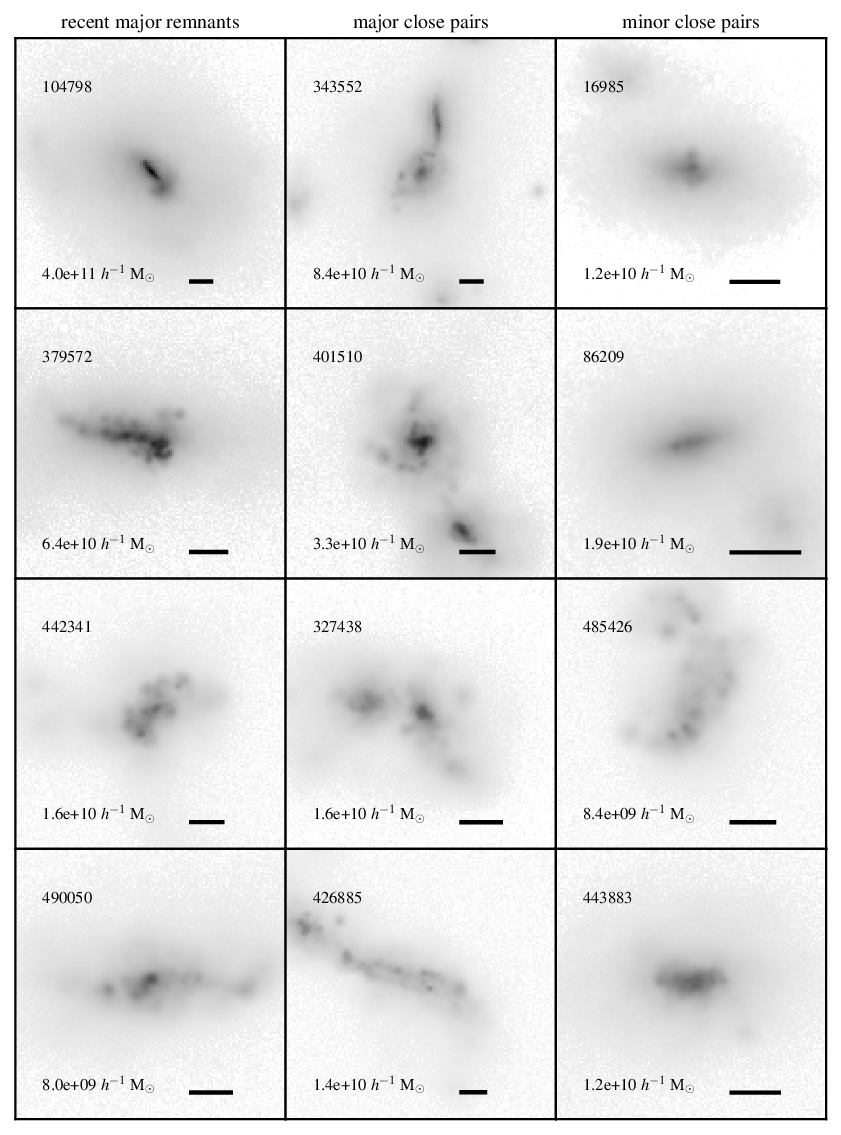}
    
    \caption{Examples of g-band mock images for major merger remnants
(left column), major close pairs (central column) and minor close pairs (right
column). The scale of the image is indicated by a 10~$h^{-1}$kpc horizontal bar. Also shown are the subhalo identification number (id) and the stellar mass of the subhalo that appear at the centre of each image.}
    \label{fig:axample_merg}
\end{figure*}

Figure~\ref{fig:axample_merg} shows examples of g-band mock images for
major merger remnants, major close pairs and minor close pairs at $z=0$.

\section{Mock images analysis}
\label{sec:image_analysis}

\subsection{Image degradation}

\begin{figure*}
    \centering
    \includegraphics[width=0.7\textwidth]{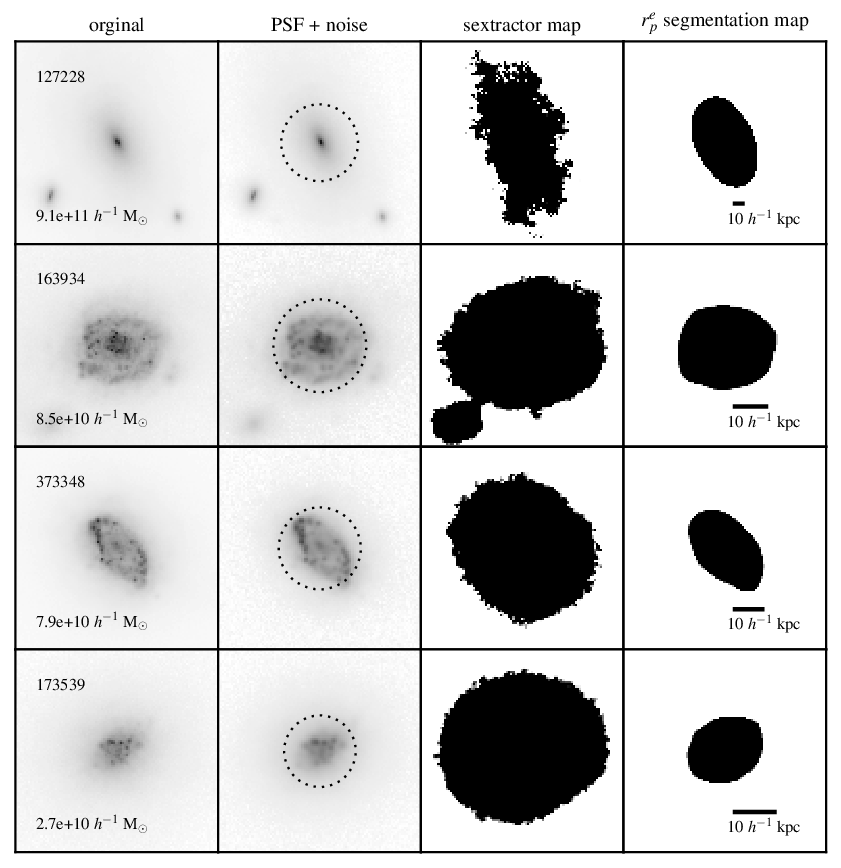}
    \caption{Example of g-band mock images before and after the image
degradation procedure and resulting segmentation maps at $z=0$, arranged by increasing
stellar mass from bottom to top. The first column shows the original images. The second
column shows the resulting images after degradation, the dashed region
represents the circular Petrosian radius. The third column shows the
segmentation maps obtained from \textsc{sextractor} and the final column displays the
segmentation maps used to compute the $G$ and $M_{20}$ statistics as described in section
\ref{sec:measurements}.}
    \label{fig:axample_seg}
\end{figure*}

Similarly to the procedure described by  \citet{snyder_galaxy_2015}, we
transform the noise-free mock images in the g-band to mimic the Sloan Digital Sky Survey
(SDSS) main galaxy sample.  Firstly, we assume that all images are at $z \sim
0.05$. Secondly, we  convolve each idealized image with a Gaussian PSF with a
full-width at half maximum (FWHM) of 1 arcsec simulating the effects of
seeing. Thirdly, we re-bin the images to a constant pixel scale of 0.24
arcsec. Finally, we add Gaussian noise to the images such that the average
\protect{signal-to-noise} ratio of each galaxy pixel is 25. Therefore, we simulate
only strongly detected galaxies with morphological measurements not affected
by noise. Figure \ref{fig:axample_seg} shows examples of mock images before
and after the degradation procedure.

\subsection{Morphology measurements}
\label{sec:measurements}

Each image is run through \textsc{sextractor} in order to produce initial
segmentation maps. We find that parameter values of
\textsc{detect\_minarea=50}, \textsc{detect\_thresh=0.6},
\textsc{deblend\_nthresh=32} and \textsc{deblend\_mincont=0.9} produce
acceptable segmentation maps that are able to correctly isolate the central
subhalo from other halo structures present in the mock images. No attempt is
made to manually correct segmentations maps due to the large sample size. The
third column of Fig.~\ref{fig:axample_seg} shows examples of segmentation maps
obtained using this procedure.

We evaluate each image background by selecting a 20 x 20 pixel square area
where no structure was detected by the corresponding segmentation map.
The background is removed from the image by subtracting the average pixel
value of the background region from all pixels. Then we assign to the central
subhalo all pixels belonging to the segmentation region that includes the central
pixel of the image. Finally, the remaining pixels belonging to any other
segmentation map are set to zero so no light from other structures besides the
central subhalo affects the morphological measurements. Below, we briefly
describe each of the morphological parameters used in this study.

The Petrosian radius $r_p$ is defined as the radius at which the ratio between the
surface brightness and  the mean surface brightness  is
equal to 0.2. For each subhalo, we compute  $r_p$ by adopting  circular
($r_p^c$) and elliptical ($r_p^e$) apertures.

The asymmetry parameter ($A$) is defined  as a measure of the fraction
of the light  in non-symmetric components \citep{abraham_galaxy_1996,conselice_asymmetry_2000}
\begin{equation}
    A = \sum_{i, j} \frac{|I(i, j) - I_{180}(i, j)|}{|I(i. j)|} - B_{180},
    \label{eq:asymmetry}
\end{equation}
where $I$ is the original image and $I_{180}$ is the image rotated by 180
degrees about a central pixel. The sum in equation~\ref{eq:asymmetry} is done over all pixels within  $1.5 r_p^c$
and the central pixel is determined by minimizing $A$. $B_{180}$ represents the
average asymmetry of the background and is computed in the sky region defined
above. 

Typical $A$ values depend on morphology type, with ellipticals having $A
\sim 0.02 \pm 0.02$ and  spirals, in the range $A \sim 0.07 - 0.2$.
Starburst galaxies such as  Ultra-Luminous Infrared
Galaxies (ULIRGs), which are often associated to major mergers, present values of $A \sim 0.32 \pm
0.19$ \citep{conselice_direct_2003}.

The concentration ($C$) measures the amount of light  within the galaxy
central region. It is defined as the ratio between the circular radii
containing 20 percent and that corresponding to 80 percent of the total galaxy
flux \citep{bershady_structural_2000}:
\begin{equation}
C = 5 \log \left(\frac{r_{80}}{r_{20}}\right).
\end{equation}
Following standard procedures \citep{conselice_direct_2003,lotz_major_2011}, we
compute the total flux  within 1.5 $r_p^c$ of the galaxy centre defined by  the pixel that minimizes $A$.

The clumpiness ($S$) quantifies the degree of small-scale structure
\citep{conselice_direct_2003} and is defined as
\begin{equation}
    S = \sum_{i, j} \frac{|I(i,j) - I_S(i,j)|}{|I(i,j)|} - B_S,
\end{equation}
where $I_S$ is the image smoothed by a 2D boxcar of width 0.25 $r_p^c$ and
$B_S$ is the average clumpiness of the background. Like $A$ and $C$, $S$ is
also summed over 1.5 $r_p^c$ but the central 0.25 $r_p^c$ region is excluded
to avoid the extremely bright galactic cores.

The gini coefficient ($G$) measures the degree of inequality in the light
distribution and is computed as
\begin{equation}
    G = \frac{1}{|\vec{X}| n (n-1)} \sum_{i}^{n}(2i - n - 1) |X_i|,
\end{equation}
where $X_i$ represent pixel values assigned to a galaxy, sorted into increasing flux order.

A $G$ coefficient of zero means that the galaxy light is evenly distributed
among all pixels, while values approaching one imply that a few pixels
concentrate most of the light. Unlike $C$, $G$ does not make any assumption
regarding the underlying morphology and is therefore sensitive to regions of
flux concentration outside the galactic centre.

Because the gini coefficient is very sensitive to which pixels are assigned to
the galaxy, we follow a prescription similar to \citet{lotz_new_2004} to
obtain additional segmentation maps that result in robust values of $G$.
Starting from the central image pixel, we compute a binary segmentation mask
employing an 8-connected structure detection algorithm. The mask is built by
accepting all pixels that are 8-connected to previously accepted pixels and
have values above a given threshold. We use as threshold the average pixel
value at a $r_p^e$ distance from the centre. The use of the elliptical
Petrosian radius is fundamental to obtain consistent results at all galaxy
orientations. The last column of Fig.~\ref{fig:axample_seg} shows examples of
segmentation maps obtained from the described procedure (see appendix
\ref{sec:appendix} for a comparison with Lotz's results).

The total second-order moment $M_{\textnormal{tot}}$ is defined as
\begin{equation}
    M_\textnormal{tot} = \sum_i^n M_i = \sum_i^n I_i ((x_i - x_c)^2 + (y_i - y_c)^2)
\end{equation}
where $I_i$ is the flux in each pixel, ($x_i$, $y_i$) represent individual
pixel coordinates and ($x_c$, $y_c$) denotes the galaxy centre. The sum is
performed over all pixels assigned to the galaxy by the same segmentation map
used to compute the $G$ index. The centre is computed by finding ($x_c$, $y_c$)
such that $M_\textnormal{tot}$ is minimized.

The second order moment of the brightest pixels of a galaxy is sensitive to
the spatial distribution of bright nuclei, spiral arms, bars and off-centre
star clusters. $M_{20}$ is defined as the normalized second order moment of
the brightest 20 per cent of a galaxy flux. To compute $M_{20}$, we sort the
galaxy pixels in descending order of flux, sum $M_i$ until the value equals 20
percent of the total galaxy flux and then normalize by $M_\textnormal{tot}$:
\begin{equation}
    M_{20} = \log_{10} \left(\frac{\sum_i M_i}{M_\textnormal{tot}} \right), \textnormal{while} \sum_i f_i < 0.2 f_\textnormal{tot}.
\end{equation}
The normalization by $M_\textnormal{tot}$ removes the dependence on total
galaxy flux. 

According to equation 6, $M_{20}$ is always a negative quantity. For normal
\protect{early-type} galaxies typical $M_{20}$ values are $\sim -2$, while
for \protect {late-type} galaxies, $M_{20} \sim -1.5$. It has been shown
that mergers present higher values, $M_{20} \geq -1$, specially those with
multiple nuclei \citep{lotz_galaxy_2008}.

\section{Analysis}

\subsection{$G-M_{20}$ criteria}

\begin{figure}
    \centering
    \includegraphics[width=1\columnwidth]{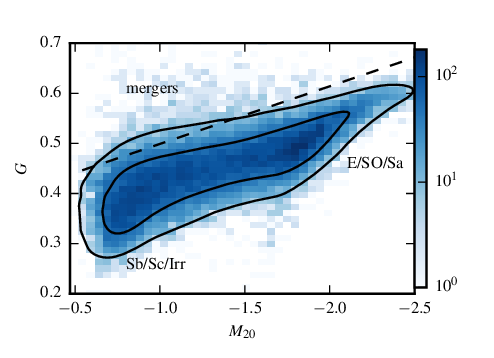}

    \caption{$G - M_{20}$ relation for the full I-1 galaxy sample. The
colorbar indicates the number of galaxies in each bin. Perturbed morphologies
are located above the LPM04 empirical demarcation line (shown as a dashed
line). Bulge-dominated galaxies are found towards the upper right corner while
disc-dominated galaxies are located  in the lower left corner. Solid lines
mark the regions which enclose 95 and 68 percent of subhalos.}

    \label{fig:all_gini_m20}
\end{figure}

In Fig.~\ref{fig:all_gini_m20} we show the $G-M_{20}$ statistics distribution
for our complete I-1 galaxy sample. Our computations reproduce
the morphological trends found by \citet{snyder_galaxy_2015}: quenched,
\protect{bulge-dominated} galaxies have large $G$ ($\sim0.6$) and low
$M_{20}$ ($\sim-2.5$) values, while \protect{disc-dominated} galaxies have lower
$G$ ($< 0.5$) and higher $M_{20}$ ($>-1$) values. We point out that the mock images
employed in the present work differ from the ones used in
\citet{snyder_galaxy_2015} in that they include all material from the FOF halo
that falls within the mock image FOV (see Section~\ref{sec:mock_observations}
for details). For individual galaxies, we find that our $G-M_{20}$ statistics generally
differ less than 10 percent from the one reported by
\citet{snyder_galaxy_2015}.

\begin{figure*}
    \includegraphics[width=0.8\textwidth]{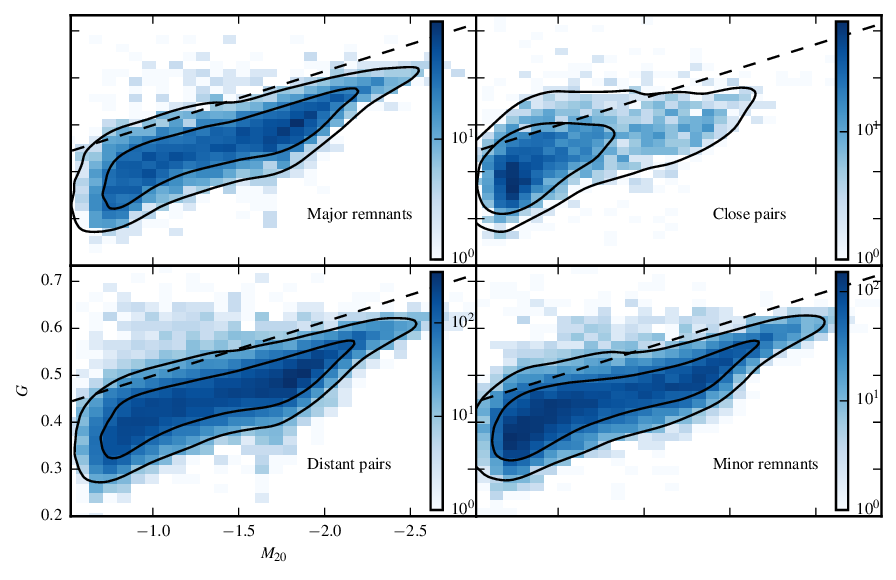}

    \caption{$G-M_{20}$ relation for galaxies that experienced a major merger
in the last 2 Gyr (upper left panel), galaxies having a close companion at a
distance d $<$ 20 $h^{-1}$ kpc (upper right panel), galaxies having a close
companion at a distance 20 $h^{-1}$ kpc  $<$ d $<$ 100 $h^{-1}$ kpc (lower
left panel) and  galaxies that experienced a minor merger in the last 2 Gyr
(lower right panel). Colorbars indicate the number of galaxies in each
bin. The dashed line separates mergers from normal galaxies according to
LPM04. Solid lines mark regions which enclose 95 and 98 percent of subhalos.}

    \label{fig:scatter_gini_m20}
\end{figure*}

In Fig.~\ref{fig:scatter_gini_m20}, we analyse the G-M$_{20}$ morphologies for
galaxies in the merger remnants and pairs subsamples selected from the
I-1 simulation. We find that the morphological indicators for major merger
remnants and close pairs differ the most from those found in the full
sample, as expected for ongoing mergers, since they present disturbed
morphologies. \citetalias{lotz_new_2004} found that visually classified mergers
in the \citet{borne_evidence_2000} observations of local ULIGRs could be separated
from \textit{normal} galaxies by 
\begin{equation}
    G > -0.115 \, M_{20} + 0.384.
    \label{eq:LPM04_criteria} \end{equation}
This relation is displayed in Fig.~\ref{fig:all_gini_m20} as well as in all
panels of Fig.~\ref{fig:scatter_gini_m20} and constitutes an empirically
derived demarcation line to separate irregular and disturbed morphologies,
often caused by mergers, from \textit{normal} unperturbed galaxies.

Following \citet{snyder_galaxy_2015}, we define the merginess as the
perpendicular distance to the LPM04 demarcation line. We assign
positive (negative) values to  points above (below) this line. The merginess
provides a qualitative estimation of the level of morphological disturbance
present. As can be seen in Fig.~\ref{fig:scatter_gini_m20}, and in
Table~\ref{tab:gini_m20_number} a significant number of close pairs have
positive merginess, indicating the presence of disturbed morphologies, while
all other samples present a lower proportion of galaxies in the merger zone.
From Table~\ref{tab:gini_m20_number} a clear hierarchy in the proportion of
galaxies with positive merginess can be found, with close pairs presenting the
highest percentage (10.1\%), follow by major mergers remnants (7.0\%), minor
merger remnants (4.2\%), and distant pairs (3.0\%). Lastly, the unperturbed
isolated galaxies sample have the lowest proportion of galaxies above the
demarcation line (1.6\%).

\begin{table}
    \centering
    \caption{Merginess frequency. Number and percentages of mock
      galaxies above and below the LPM04 empirical merger demarcation
      line for merger remnants, galaxy pairs and unperturbed galaxies. Each class
    has been normalized to the total number of members in the subsample. All four cameras are included.}
    \begin{tabular}{lrrrr}
        \hline
        Class & \multicolumn{2}{c}{Merginess $\geq$0} & \multicolumn{2}{c}{Merginess $<$ 0}\\
        & N & Percentage & N & Percentage \\
        \hline
        Close pairs & 300 & 10.1 & 2676 & 89.1 \\
        Major mergers & 90 & 7.0 & 1198 & 93.0 \\
        Minor mergers & 635 & 4.2 & 14437 & 95.8 \\
        Distant pairs & 607 & 3.0 & 19577 & 97.0 \\
        Unperturbed & 38 & 1.6 & 2290 & 98.4 \\
        galaxies       & & & & \\
        \hline
    \end{tabular}
    \label{tab:gini_m20_number}
\end{table}

Close pairs appear clustered at M$_{20} >$ -1 consistent with the detection of
multiple nuclei within the segmentation map and comparable to the values found
by  \citet{lotz_new_2004} for double and multiple nuclei ULIRGs, which present
higher (M$_{20} \sim -1$) values that single nuclei ULIRGs (M$_{20} \sim -2$).

\subsection{Merger remnants}

According to \citet{lotz_effect_2010-1}, $G - M_{20}$ morphologies are
particularly sensitive to mergers with baryonic mass ratios between 1:1 and
10:1, during time-scales lasting 0.2-0.4 Gyr. The disturbed morphologies are
more noticeable during the close approaches and the final merger stages. They
also find that major merger remnants observed after more than 1 Gyr of the
event present morphologies similar to  early-type spirals while minor mergers
are found to have minimal effects on the $G$ and $M_{20}$ values of their
remnants.

\begin{figure} \centering
    \includegraphics[]{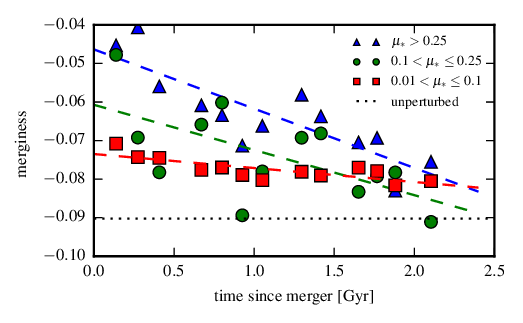}
    \caption{ 
    Mean merginess as a function of elapsing time since  the last 
    major merger ($\mu_* > 0.25$; blue triangles), minor mergers ($0.1
    < \mu_* \leq 0.25$; green circles)
    and very minor mergers  ($0.01 < \mu_* \leq 0.1$;  red squares). The dashed lines
    represent linear regressions fitting  to the corresponding
    data. For comparison we included the mean merginess of the unperturbed sample (horizontal dotted line).
    }
    \label{fig:mean_merginess_in_time}
\end{figure}

In Fig.~\ref{fig:mean_merginess_in_time} we show the mean merginess of
galaxies as a function of time since the last merger event. We find that the
morphological disturbance is larger for galaxies having just experienced a
major merger (blue triangles). A very good correlation signal is found between
the merginess and the elapsing time since the last major merger. Events
occurring  more than 2 Gyr ago show mean merginess comparable to the average
value of the unperturbed sample of galaxies. Spite of the fact that the mean
merginess remains negative for all time, the good correlation signal shows
that it is still possible to statistically classify recent mergers by
using the $G-M_{20}$ statistic.

The mean merginess for minor mergers ($0.1 < \mu_* \leq 0.25$) is also shown
in Fig.~\ref{fig:mean_merginess_in_time} as green circles. This subsample of
galaxies is noisier and has lower values for a given elapsing time to that of
major mergers, but shows a similar correlation with time, indicating that the
$G-M_{20}$ is still sensitive to mergers in this mass ratio range. In contrast, the mean
merginess for very minor mergers ($0.01 < \mu_* \leq 0.1$) shows a much weaker
dependence with time, corroborating the result found by \citet{lotz_galaxy_2008} which
suggested that very minor mergers do no significantly affect the final values
of $G$ an $M_{20}$ of their remnants. The dashed lines in
Fig.~\ref{fig:mean_merginess_in_time} represent linear regression fits to
the corresponding mean merginess. The parameters of the fitting are summarized
in Table~\ref{tab:merginess_in_time_fit}.

\begin{table}
    \centering
    
    \caption{Parameters of the linear regression fits of the mean    merginess
as a function of the elapsing time to the last merger       event for galaxy
mergers with different mass ratios $\mu_*$ (Fig.~\ref{fig:mean_merginess_in_time}).}

    \begin{tabular}{lcc}
        \hline
        Mass ratio $\mu_*$  & a   & b  \\
        &  dex Gyr$^{-1}$  &  dex \\
        \hline
        $\mu_* \geq  0.25$  & $-0.015 \pm 0.003$ & $-0.04$ \\
        $0.25 < \mu_* \leq 0.1$ & $-0.012 \pm 0.003$ & $-0.006$ \\
        $0.1 < \mu_*  \leq 0.01$ & $-0.004 \pm 0.001$ & $-0.07$ \\
        \hline
    \end{tabular}
    \label{tab:merginess_in_time_fit}
\end{table}

\begin{figure*}
    \includegraphics[]{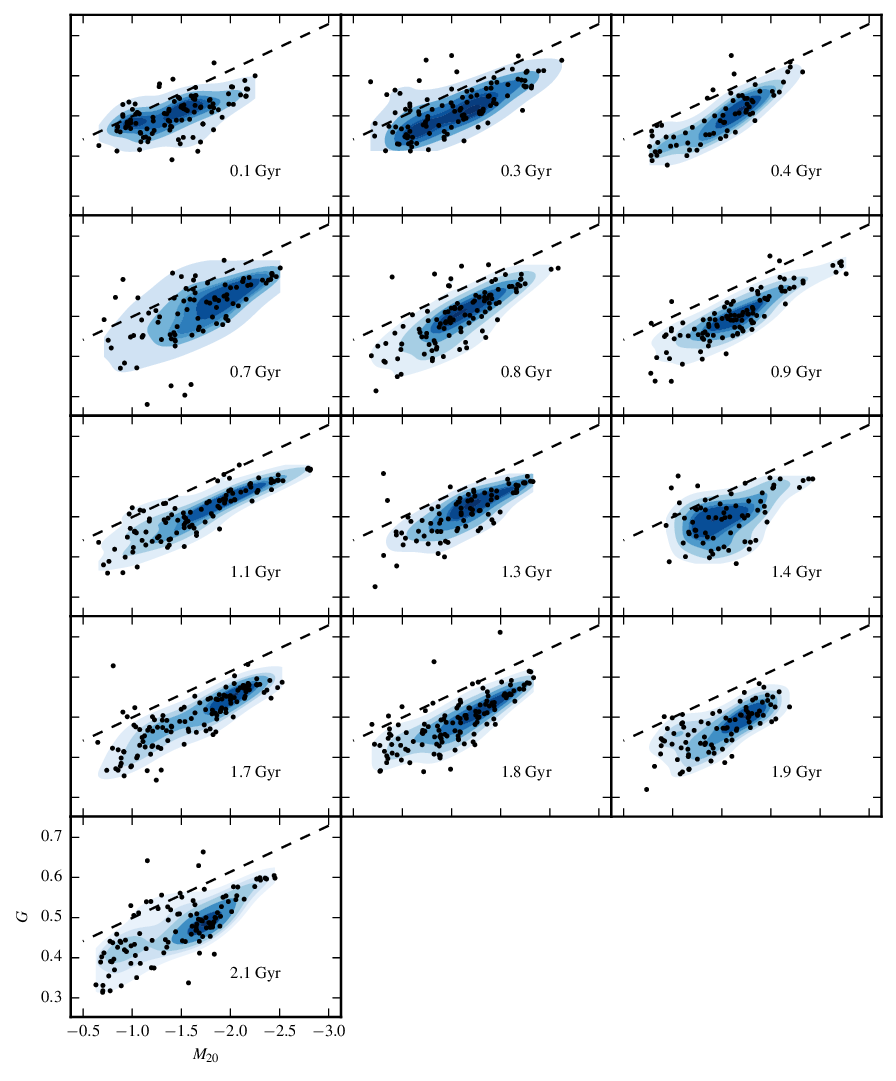}
    
    \caption{$G - M_{20}$ morphologies for galaxies having experienced
the last major merger event at  increasing times for each panel. Points
represent the major merger remnants, while the shaded contours denote regions
that enclose 90, 70, 50, 30 and 20 per cent of remnants (from light blue to
dark blue colours) . The dashed line corresponds to the LMP04 demarcation
line.}

    \label{fig:majors_in_time}
\end{figure*}

To visualize how the merger remnants are located in $G - M_{20}$ plane
according to the elapsing time to the last merger event, we displayed
them in Fig.~\ref{fig:majors_in_time}.  As expected from our previous
discussion,  recent major
mergers are more likely to be found above the empirical demarcation line than
older merger remnants. This is also in agreement with \citet{lotz_effect_2010-1}. findings
which indicate that more recent merger remnants are more likely to be
classified positively.  However, we note that  a large number of major merger
remnants are located below the demarcation line even for very recent merger.
This is consistent with previous findings suggesting that the $G-M_{20}$
method yields an  incomplete classification \citep{kampczyk_simulating_2007}.

Interestingly, we also find that there is a shift in the position of merger
remnants with time along the relation, such that  very recent mergers appear
to be clustered around $M_{20} \sim -1.5$ while older remnants are shifted
towards lower values ($M_{20} \sim -2$). As mentioned before, higher $M_{20}$
values are an indication of multiple nuclei detected within a segmentation map
and are expected to appear in the time immediately after the final encounter.
At intermediate times (t $\sim 1$ Gyr) a tail towards bulge dominated galaxies
can be found. While, at late stages (t $\sim 2$ Gyr) major remnants are found
in the zone between late type galaxies and bulge dominated early type galaxies
indicating that despite the major merger, many galaxies manage to retain or
recover their disc structures. This result agrees well with previous
findings by \citet{robertson_mergerdriven_2006} that gas-rich mergers can form
rotationally supported gaseous structures from residual angular momentum after
the final coalescence, with similar trends found in zoom-in merger simulation
by \citet{snyder_diverse_2015} and with results found by
\citet{lotz_galaxy_2008} where equal mass, gas rich isolated merger
simulations appear disc-like when observed t $> 1$ Gyr after the final
coalescence.

We next determine the effectiveness of the \citetalias{lotz_new_2004}
empirical demarcation criteria to identify major merger remnants.

For a given selection criteria such as the \citetalias{lotz_new_2004}
empirical demarcation line, it is necessary to ascertain how effectively it
selects a given subsample of galaxies. In a similar way to the method used by
\citet{huertas-company_measuring_2014} to calibrate automatic proxies of
galaxy morphology we proceed to define the following quantities:

\begin{enumerate}
    \item True positives (TP): Number of galaxies selected by the criteria which belong to the subsample.
    \item True negatives (TN): Number of galaxies not selected by the criteria which do not belong to the subsample.
    \item False positives (FP): Number of galaxies selected by the criteria which do not belong to the subsample.
    \item False negatives (FN): Number of galaxies not selected by the criteria which belong to the subsample.
\end{enumerate}
We define two additional quantities: the purity (P) and the completeness (Cp).
Purity is the percentage of selected galaxies that belong to the subsample among all galaxies selected by the criteria.
\begin{equation}
    \text{P} = 100 \times \frac{TP}{FP + TP}.
    \label{eq:purity}
\end{equation}
It effectively measures the level of contamination. For example, if 90 per cent of galaxies with positive merginess have in fact not experienced a major merger, then the purity of the \citetalias{lotz_new_2004} criteria in selecting major merger remnants will be 10 per cent.

The completeness (Cp) is the percentage of selected galaxies which belong to the subsample
among all galaxies belonging to the subsample.
\begin{equation}
    \text{Cp} = 100 \times  \frac{TP}{FN + TP}.
    \label{eq:completness}
\end{equation}
For example, if all major merger remnants have positive merginess, then the completeness of the \citetalias{lotz_new_2004} criteria in selecting major merger remnants will be 100 per cent.

We performed a visual inspection of the galaxies presenting positive
merginess in the first of the four cameras in order to determine the possible
presence of sources of contamination that would affect the computations of P
and Cp. From the original 266 galaxies we find that 4 (1.5\%) galaxies are
edge-on, a common source of confusion for the G-M$_{20}$ criteria, 43 (16.2\%)
present ring-like structures of recent star formation previously reported by
\citet{snyder_galaxy_2015,torrey_synthetic_2015}, finally we found 50 (18.8\%)
galaxies with irregular and starbursting appearance, but no sign of recent or
current interaction such as tidal tails or a close companion. These galaxies
have low stellar mass ($M_* \sim10^{10}$ M$_\odot$) and are likely the result of
stochastic recent star formation which combined to the limited numerical
resolution produced artificially large star formation regions. We also find
similar trends in the rest of the cameras. We define a clean sample of
perturbed galaxies as the galaxies that present positive merginess in the
first camera once we remove the above sources of contamination.

\begin{figure} \centering
    \includegraphics[]{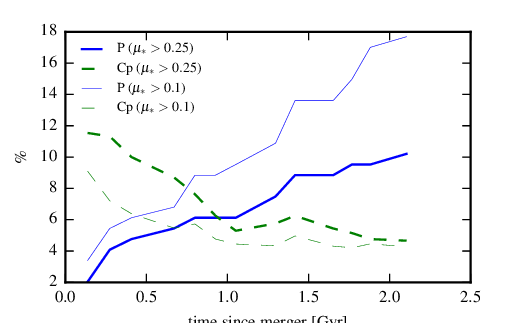}
    
    \caption{Purity (solid line) and completeness (dashed line) for merger remnants as
        a function of elapsing time since the last merger event. Thick lines
        denote remnants with $\mu_* > 0.25$, while thinner lines denote remnants
        with $\mu_* > 0.1$.}

    \label{fig:purity_matrix_majors}
\end{figure}

In Fig.~\ref{fig:purity_matrix_majors} we show the dependence of the Cp and P
quantities on the elapsing time since the last merger event using the clean
sample defined above. From the completeness we find that about 12 per cent of
major mergers occurring since  $t<0.14$ Gyr are classified as perturbed
according to the LPM04 empirical criterium. This percentage decreases towards
5 per cent at $t\sim1$ Gyr where it remains approximately constant afterwards.
This result further corroborates that this empirical criteria is sensitive  to
major mergers occurring less than 1 Gyr ago and specially sensitive to very
recent events. The purity indicates a high level of contamination for all
elapsing times, which is expected because at a given time, only a small
fraction of perturbed galaxies are expected to be a major merger remnant. We
can conclude that  $\sim 5$ per cent of galaxies with positive merginess can
be explained as a major merger remnant.
Figure~\ref{fig:purity_matrix_majors} also shows Cp and P for merger
remnants with $\mu_* > 0.1$, completeness values present a similar dependence
with the elapsing time since the merger event compared to major mergers. After
1 Gyr completeness levels off showing that the $G-M_{20}$ criteria is still
sensitive to $\mu_* > 0.1$ younger than 1 Gyr.

\subsection{On-going mergers}
\label{sec:ongoing}

\begin{figure} \centering
    \includegraphics[]{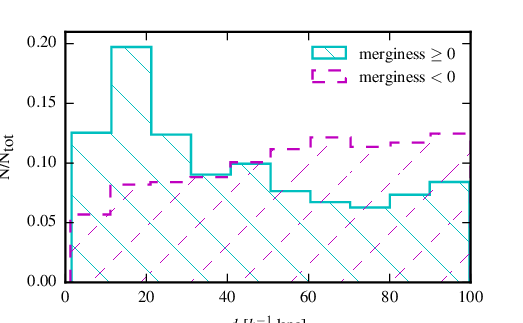}
    \caption{Distribution of relative distances between galaxies in pairs with positive merginess (cyan solid lines) 
        and negative merginess (magenta dashed line).}
    \label{fig:pair_dist_distributions}
\end{figure}

While our merger sample selected by using the merger trees represents the
remnants of mergers, our  pair sample  represents the population of on-going
interactions. In Fig.~\ref{fig:pair_dist_distributions} we show the
distribution of relative distances between pairs for those galaxies with
positive and negative merginess. There is a clear  excess of close
pairs with perturbed  morphologies. As it has been pointed out in previous
works, the largest morphological changes and star formation excess are,
detected for galaxies within $\sim 35$ h$^{-1}$ kpc  in both observational
\citep[e.g.][]{barton_tidally_2000,lambas_galaxy_2003,ellison_galaxy_2008, scudder_galaxy_2012} and
numerical \citep[e.g.][]{perez_galaxy_2006, di_matteo_star_2007} studies.

\begin{figure} \centering
    \includegraphics[]{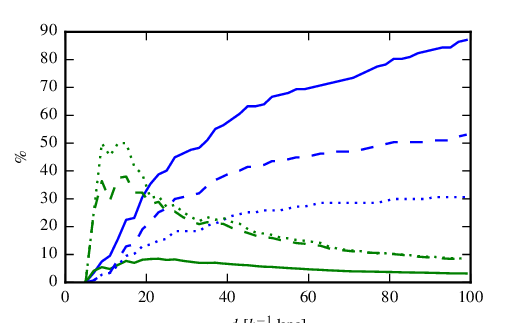}
    \caption{Purity (blue lines) and completeness (green lines) for pairs as a
function of relative distance between galaxy members when considering the
following mass ratios: $\mu_*>0.25$ (dotted lines), $\mu_*>0.1$ (dashed lines),
$\mu_*>0.001$ (solid lines)}
    \label{fig:purity_distance_clean}
\end{figure}

In Fig.~\ref{fig:purity_distance_clean} we show the purity and completeness of
the cleaned sample as a function of  distance between pair members.
When considering all mass ratios ($\mu_* > 0.001$), we find low completeness
values ($\sim5\%$) at all separations. Conversely, for $\mu_* > 0.1$ we find
completeness values that reach $\sim40\%$ at $d < 20$ $h^{-1}$ kpc . This confirms that
the $G-M_{20}$ criterium is sensitive to mass ratios larger than 0.1. We
notice that completeness increases by a factor of 1.3 for major mergers with
$d < 20$ $h^{-1}$ kpc, supporting the claim that $G-M_{20}$ is more sensitive to this kind of events.
 Apart from this, the behaviour is very similar to that determined by
imposing a limit at $\mu_* = 0.1$ as reported by \citet{lotz_effect_2010-1}.

Purity for pairs with $\mu_* > 0.1$ plateaus at $\sim50\%$ for separations greater than
$\sim 100$ $h^{-1}$ kpc. We adopt this percentage as the fiducial value for contamination of
our clean sample. This value is of fundamental importance for our derivation of
the merger rate.

\subsection{Asymmetry criteria}

The Asymmetry parameter is also commonly used to classify merger candidates.
The calibration for local mergers by \citet{conselice_relationship_2003} finds
the following merger criterium

\begin{equation}
    A \geq 0.35.
    \label{eq:conselice_criteria}
\end{equation}

\begin{figure*}
    \centering
    \includegraphics[width=0.8\textwidth]{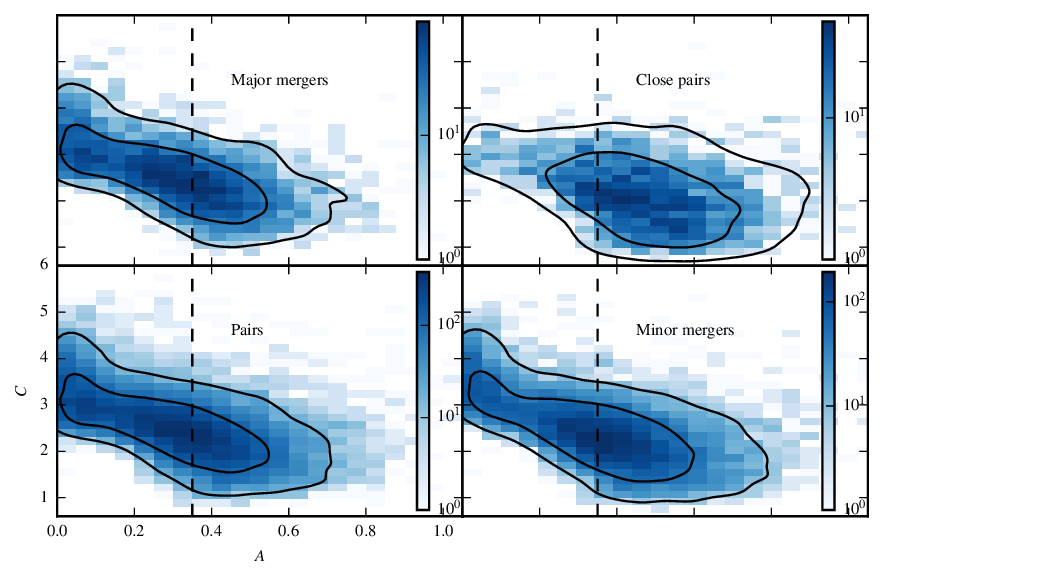}
    \caption{$A - C$ relation for  galaxies that experienced a major
      merger in the last 2 Gyr (upper left panel), for galaxies having
      a close companion at a distance $d< 20$ $h^{-1}$ kpc (upper right panel),
      for galaxies having a close companion at a distance  $d<100$ $h^{-1}$ kpc
      (lower left panel) and for : galaxies that experienced a minor
      merger in the last 2 Gyr (lower right panel) in the I-1 simulation at $z
= 0$. The
vertical line separates mergers from normal galaxies according to empirical
criteria of \citet{conselice_relationship_2003}. The contours mark
    regions that enclose 90, 70, 50, 30 and 20 percent of subhalos
    (from dark to light blue, respectively).}
    \label{fig:scatter_asymmetry}
\end{figure*}

In Fig.~\ref{fig:scatter_asymmetry} we show concentration vs. asymmetry for
galaxies in the I-1 simulation at $z = 0$. As before we treat each of the four
cameras an independent measurement. We discriminate again between major
mergers, minor mergers close pairs, and distant pairs. We find that the
condition $A \geq 0.35$ roughly divides in half the galaxy populations considered
with the exception of close pairs which present a significant excess
of asymmetry.
From this figure, then, it is clear that close pairs are those better
classified by the asymmetry criterium.

\begin{table}
    \centering
    \caption{Number and percentages of galaxies classified according
      to the criteria proposed by \citet{conselice_relationship_2003}.}
    \begin{tabular}{lllll}
        \hline
        Class & \multicolumn{2}{c}{$A \geq 0.35$} & \multicolumn{2}{c}{$A < 0.35$}\\
        & N & Percentage & N & Percentage \\
        \hline
        Major mergers & 580 & 45 & 708 & 55 \\
        Close pairs & 2356 & 79.2 & 620 & 20.8 \\
        Minor mergers & 7721 & 51.2 & 7351 & 48.8 \\
        Pairs & 8468 & 42 & 11716 & 58 \\
        Unperturbed & 955 & 41 & 1373 & 59 \\
        galaxies & & & & \\
        \hline
    \end{tabular}
    \label{tab:asymmetry_numbers}
\end{table}

\begin{figure} \centering
    \includegraphics[]{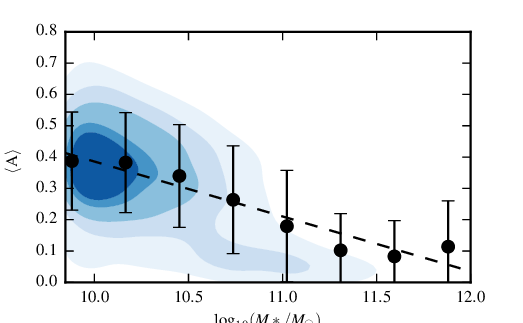}
 \caption{Mean asymmetry $A$ as a function of stellar mass. The contours
mark regions that enclose 90, 70, 50, 30 and 20 percent of
subhalos. The black dots
show the mean asymmetry in stellar mass bins, with error bars representing the
standard dispersion. The dashed line represents a linear fit to the points with slope
$-0.17\pm0.02$ and intercept $2.14\pm0.01$.}
\label{fig:mass_asymetry}
\end{figure}

As can be seen in Fig.~\ref{fig:mass_asymetry}, the mean asymmetry of galaxies
increases with decreasing mass, resulting in a large fraction of subhalos with
stellar masses 10$^{10}$ M$_\odot$ having $A \geq 0.35$. As mentioned
before, galaxies with masses around 10$^{10}$ M$_\odot$ show a
trend towards having more irregular morphologies due to stochastic star
formation and limitations of the numerical resolution of the simulation.
Hence, in order to better assess the behaviour of $A$ for galaxy pairs, the
subsample is divided in mass intervals.

\begin{figure} \centering
    \includegraphics[]{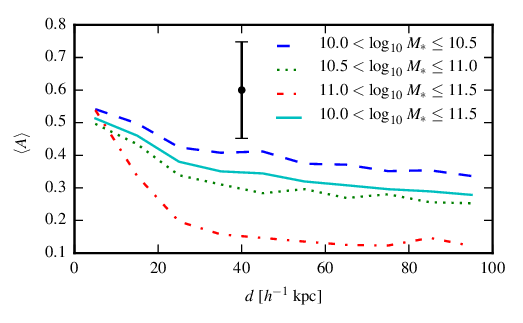}
    \caption{Mean asymmetry as a function of the relative distance to a  companion galaxy
        for different stellar mass intervals: $10 < log_{10} M_{*} \leq 10.5$ (blue dashed line), $10.5 <
        log_{10} M_{*} \leq 11$ (green dotted line) and $11 < log_{10} M_{*} \leq 11.5$ (red dashed-dotted line). The solid cyan line represents the full mass range $10 < log_{10} M_{*} < 11.5$. The data point with error bars indicate the typical dispersion at a given pair separation.}
    \label{fig:mean_asymmetry_distance}
\end{figure}

Regardless of stellar mass, the mean asymmetry of subhalos increases for close
pairs, as can be seen in Fig.~\ref{fig:mean_asymmetry_distance} where we show
$\left<A\right>$ as a function of distance to the closest companion galaxy for mass bins in the
ranges $11 < \log_{10} M_{*} \leq 11.5$, $10.5 < \log_{10} M_{*} \leq 11$ and $10 < \log_{10}
M_{*} \leq 10.5$. We also include the relation for the whole sample ($10
< \log_{10} M_{*} \leq 11.5$). All mass bins show an increase of the $\left<A\right>$ over the reference
value at $d \sim 30$ $h^{-1}$ kpc

The complete sample shows an increase of the $\left<A\right>$ over the reference value
at $d \sim 30$ $h^{-1}$ kpc. Galaxies with $\log_{10} M_{*} > 11.0 $ show the sharper increases but for galaxies at lower
separation, $d \sim 20$ $h^{-1}$ kpc and for larger pair separations $\left<A\right>$ is
lower than 0.2. As we take smaller galaxies, the $\left<A\right>$ are higher and
show a weaker variation  with the relative distance between pairs. However,
smaller galaxies get to values larger than $\left<A\right> \sim 0.35$ at larger pair
separations. The smaller mass interval might be more affected by resolution
problems which produce spurious signals of disturbances in the non-parametric
morphologies. However, the trend is present progressively as one moves from
higher mass to lower mass.

\begin{figure} \centering
    \includegraphics[]{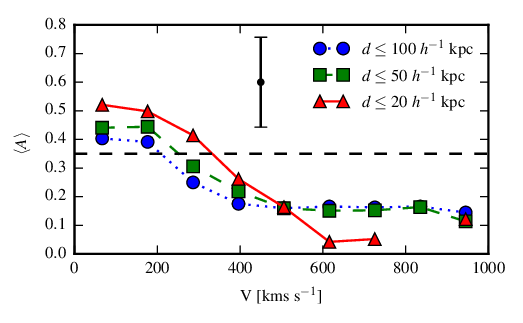}
    
    \caption{Mean asymmetry as a function of relative velocity between
galaxy pairs for three pair separation. The horizontal       dashed line shows
the reference value  $A=0.35$. The data point with error bars indicate the
typical mean standard deviation. This trend shows the very weak effects
of flybys in triggering a morphology perturbation.}

    \label{fig:mean_asymmetry_velocity}
\end{figure}

In order to explore further if the increase of asymmetry in close pairs is
produced by physical disturbances in the galaxies such as tidal tails or induced star
formation and not by light contamination from the secondary galaxies, we study the
asymmetry as a function of relative velocity between pair members.
This analysis will also allow us to detect the role play by those
pairs with larger velocity separations which  have a higher
probability to be flyby events. 
As can be seen in Fig.~\ref{fig:mean_asymmetry_velocity}, there is an increase in
mean asymmetry over the reference value for close pairs having a relative velocity smaller that
$\sim 400$ km s$^{-1}$ 
as is expected for interacting pairs \citep[e.g][]{lambas_galaxy_2003} For
larger relative velocities, the asymmetry gets to value of around $\left<A\right>
\sim 0.1-0.2$. As galaxy pairs with larger separations are
incorporated, the signal  of anti-correlation decreases.  
However, the subsample of closer galaxy pairs clearly show the
correlation between disturbances and mergers and suggests that
flyby events will not be able to produce such a significant impact, on average.

\begin{figure} \centering
    \includegraphics[]{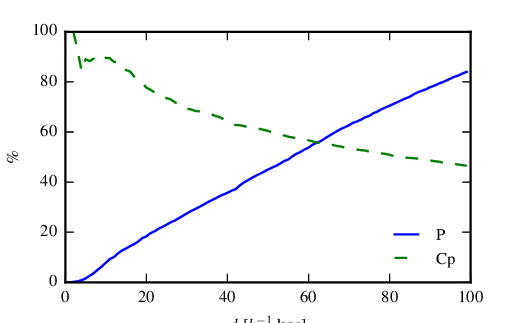}
    \caption{$A - C$ purity (solid line) and completeness (dashed line) for
for galaxy pairs as a function of pair separation.}
    \label{fig:purity_asymmetry_distance}
\end{figure}

Fig.~\ref{fig:purity_asymmetry_distance} shows the dependence of purity and
completeness on pair separation for the asymmetry merger criterium. We find that
the completeness increases from 60 per cent  for  $d < 60$ $h^{-1}$ kpc  to 90
per cent  within  $d < 10$ $h^{-1}$ kpc.

\section{Galaxy merger rate}

Following \citet{lotz_major_2011} we define the volume-averaged galaxy merger
rate $\Gamma_\textnormal{merg}$ as the number of ongoing merger events per
unit comoving volume ($\phi_\textnormal{merg}$), divided by the time
$T_\textnormal{merg}$ between the initial encounter and the final merger
stage.

\begin{equation}
    \Gamma_\textnormal{merg} = \frac{\phi_\textnormal{merg}}{T_\textnormal{merg}}
    \label{eq:merger_rate}
\end{equation}

The number density of galaxies classified as galaxy mergers depends on the
average time-scales ($\left<T_\textnormal{obs}\right>$) during which the
galaxy can be identified by some of the morphological methods discussed above,
such that

\begin{equation}
    \phi_\textnormal{merg}' = \phi_\textnormal{merg} \frac {\left<T_\textnormal{obs}\right>}{T_\textnormal{merg}}.
    \label{eq:identified_merger}
\end{equation}
The galaxy merger rate can be calculated from the observed number density of galaxy merger candidates as
\begin{equation}
    \Gamma_\textnormal{merg} = \frac {\phi_\textnormal{merg}'}{\left<T_\textnormal{obs}\right>}.
    \label{eq:identified_merger_rate}
\end{equation}

Instead of $\phi_\textnormal{merg}$, many authors estimate the fractional merger
rate $R_\textnormal{merg}$ defined as

\begin{equation}
    R_\textnormal{merg} = \frac{f_\textnormal{merg}}{\left<T_\textnormal{obs}\right>},
    \label{eq:fraction merger}
\end{equation}

where $f_\textnormal{merg}$ is the fraction of galaxies identified as mergers
for a given galaxy sample. We can relate $\phi_\textnormal{merg}'$ to
$f_\textnormal{merg}$ by using:

\begin{equation}
    \phi_\textnormal{merg}' = f_\textnormal{merg} n_\textnormal{gal},
    \label{eq:observed_number_to_fractionmerger}
\end{equation}
where $n_\textnormal{gal}$ is comoving number density of galaxies.

A correction factor can be applied to the merger fraction to account for
contamination from objects that are not mergers, such that
\begin{equation}
    f_\textnormal{merg} = C_\textnormal{merg} f_\textnormal{merg}^\textnormal{obs},
    \label{eq:corrected_fraction}
\end{equation}
where $f_\textnormal{merg}^\textnormal{obs}$ is the fraction of galaxies
identified as mergers before the correction is applied.

\subsection{Average merger observability time-scale}
\label{sec:time-scale}

Individual observability time-scales $T_\textnormal{obs}$ were calculated by
\citet{lotz_galaxy_2008,lotz_effect_2010,lotz_effect_2010-1} for a suite of
N-body/SPH isolated merger simulations spanning a range of galaxy masses, mass
ratios, gas fractions, orientations and orbital parameters. SDSS-g mock images
were used to calculate the time during which particular merger simulations
would be counted as perturbed according to the $G-M_{20}$ criterium. They
found that observability time-scales depend mostly on the mass ratio and gas
fraction of galaxies involved in the merger, while orientation, orbital
parameters and the final merger mass had little impact on
$T_\textnormal{obs}$.

Following \citet{lotz_major_2011} we compute the average observability time-scale
$\left<T_\textnormal{obs}\right>$ expected for the I-1 simulation at $z=0$ as:

\begin{equation}
 \left<T_\textnormal{obs}\right> = \sum_{i, j} w_{i, j} \times T_{i, j}
 \label{eq:average_obs}
\end{equation}

where $w_{i, j}$ is the fraction of mergers at $z=0$ with stellar mass ratio
$i$ and baryonic gas fraction $j$, and $T_{i, j}$ is the observability
time-scale corresponding to mergers with stellar mass ratio
$i$ and baryonic gas fraction $j$. 

\begin{figure} \centering
    \includegraphics[]{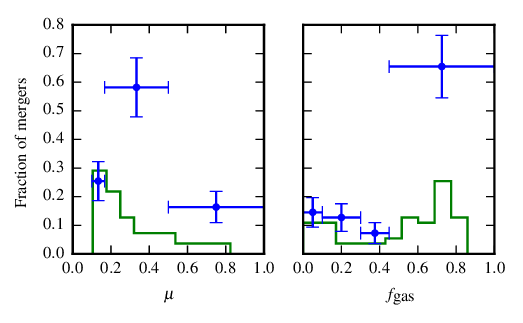}
    \caption{ Normalized distribution of stellar mass ratios (left panel) and gas
      fractions (right panel) for
      mergers in the I-1 simulation at $z=0$ (green lines).
      Galaxies were selected with
      total stellar mass M$_{*}$ $>$ $10^{10}$ M$_\odot$ and stellar
      mass ratios $\mu_* > 0.1$. For comparison, the estimations by
      using the intervals of \citet{lotz_major_2011} are included. Error bars  in the y-axis denote
      Poisson noise in the number of mergers in a given interval,
      while error bars in the x-axis denote bin widths.}
    \label{fig:fraction_gas_ratio}
\end{figure}

Figure~\ref{fig:fraction_gas_ratio} shows the normalize distribution of
stellar mass ratios and gas fractions for I-1 mergers at $z=0$. We
also estimated them for the same parameter space explored by the isolated
merger simulations detailed above: $1-1/2$, $1/2-1/6$ and $1/6-1/10$ $\mu_*$ intervals, and
$0.0-0.1$, $0.1-0.3$, $0.3-0.45$ and $0.45-1.0$ baryonic gas fraction
($f_\textnormal{gas}$) intervals.

Since for the $G-M_{20}$ criterium the individual observability time-scales
$T_{i,j}$ are not a strong function of $f_\textnormal{gas}$
\citep{lotz_effect_2010}, we average $T_{i,j}$ over $f_\textnormal{gas}$ for
each of the three mass bins. Then, we perform the sum in
Eq.\ref{eq:average_obs} over the stellar mass ratios ($i$). Merging the
$f_\textnormal{gas}$ bins also contributes to improve the statistics in $w_{i, j}$,
otherwise some bins remain with a low number of mergers (less than 5).

We obtained a value of $\left<T_\textnormal{obs}\right> \sim 0.20$ Gyr for the
I-1 simulation at $z=0$. \citet{lotz_major_2011} estimated the same
cosmological average observability value using three different cosmological
galaxy evolution models: 0.2 Gyr \citep{somerville_semi-analytic_2008}, 0.21
Gyr \citep{croton_many_2006} and 0.22 Gyr \citep{stewart_galaxy_2009}. It is
encouraging that similar $\left<T_\textnormal{obs}\right>$ are obtained by using different theoretical
approaches: the semi-analytic approach \citep{croton_many_2006,somerville_semi-analytic_2008}, halo abundance matching
\citep{stewart_galaxy_2009} and N-body hydrodynamical cosmological simulations
such as I-1.

\subsection{Intrinsic and morphological merger rates}

Taking advantage that the simulation provides us with the assembly history of
galaxies via the merger trees, we can calculate the intrinsic merger rate as a
function of minimal mass ratio $\mu_\textnormal{min}$ as

\begin{equation}
    R_\textnormal{merg}^\textnormal{intr.} (\mu_\textnormal{min}) = \frac{N(\mu_\textnormal{min})}{T_\textnormal{max}}, 
    \label{eq:intrinsic_rate}
\end{equation}
where $T_\textnormal{max} = 0.13$ Gyr is the elapsing time between the $z=0$ snapshot and the previous snapshot and $N(\mu_\textnormal{min})$ is the number of galaxies having experience a merger with mass ratio larger than $\mu_\textnormal{min}$.

\begin{figure} \centering
    \includegraphics[]{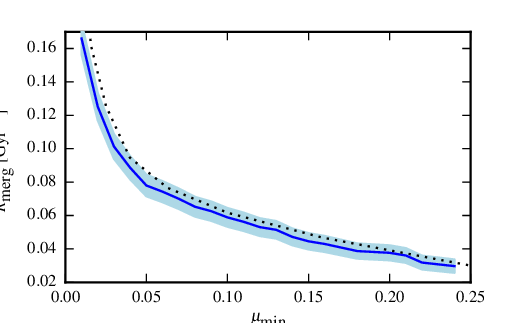}
    \caption{The cumulative intrinsic merger rate per galaxy for the
      I-1 simulation at $z=0$ (solid black line). The blue shaded
      region represents the Poisson noise. For the sake of comparison,
       the  cumulative merger rate derived integrating the
      \citet{rodriguez-gomez_merger_2015} fitting function for the I-1
      galaxy-galaxy merger rate is included (dotted line).}
    \label{fig:illustris_rate}
\end{figure}

 Figure~\ref{fig:illustris_rate} shows the cumulative (with
 respect to mass ratio) intrinsic merger ratio for the I-1 simulation at $z=0$. From this
 figure a quick assessment of the merger rate is possible. For
 example, the intrinsic rate at $\mu_* > 0.1$ is approximately 0.06
 Gyr$^{-1}$, which implies that roughly one in every 17 galaxies has experienced a  $\mu_* > 0.1$ merger in the last Gyr. Similarly, one in every 33 galaxies has experienced a major merger in the last Gyr.

We  compare the intrinsic rate to the rate which would be derived
using  the non-parametric morphologies as proposed by \citet{lotz_major_2011}
\begin{equation}
    R_\textnormal{merg}^\textnormal{morph.} (\mu_\textnormal{min}) = 0.5 \frac{N_\textnormal{above}}{\left<T_\textnormal{obs}\right>}, 
    \label{eq:lotz_rate}
\end{equation}
where $N_\textnormal{above}$ is the number of galaxies which are found above
the LPM04 demarcation line. The value of 0.5 is the fiducial
$C_\textnormal{merg}$ that corresponds to the level of contamination derived
in Section~\ref{sec:ongoing} According to the results of
Section~\ref{sec:time-scale}, we adopt $\left<T_\textnormal{obs}\right> = 0.2$
Gyr.

Therefore, by using our analysis of the I-1 simulation,  we can compare the intrinsic merger rate with the morphologically derived one
\begin{equation}
    R_\textnormal{merg}^\textnormal{intr.} (\mu_\textnormal{min}) = C' R_\textnormal{merg}^\textnormal{morph.}.
    \label{eq:correction_value}
\end{equation}
The $C'$ factor can be interpreted as a correction factor that brings the
morphologically derived rate to the intrinsic merger rate of the simulation.
If the computed $\left<T_\textnormal{obs}\right>$ is correct, that is, if the
isolated merger observability time-scales $T_{i, j}$ are correct, then one
would expect the $C'$ factor to be close to unity.

\begin{figure} \centering
    \includegraphics[]{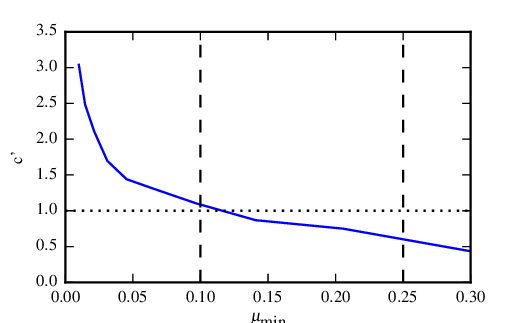}
    \caption{Correction fraction $C'$ defined in
      equation~\ref{eq:correction_value} as a function of minimal mass
      ratio $\mu_\textnormal{min}$. The horizontal dotted line at
      $C'=1$ intersects the curve at the mass ratio for which the
      morphologically derived rate recovers the intrinsic rate of
      I-1. As can be seen the intersection happens close to $\mu_* =
      0.1$, which is consistent  with the ratio for which $G-M_{20}$
      becomes sensitive to perturbations. The vertical dashed lines
      are shown for reference only;  they denote the positions at $\mu_* = 0.1$ and $\mu_* = 0.25$}
    \label{fig:correction_fraction}
\end{figure}

Fig.~\ref{fig:correction_fraction} shows the correction factor $C'$ as a
function of $\mu_\textnormal{min}$. We find that for a minimum mass ratio of
$\sim0.1$ no further correction factor is necessary to recover the intrinsic
merger rate of the I-1 simulation. \citet{lotz_major_2011} found that the
averaged observability time-scales derived from isolated pair simulations
resulted in a global merger rate  around a order of magnitude larger than
predicted theoretical values \citep{somerville_semi-analytic_2008,stewart_galaxy_2009,croton_many_2006,hopkins_mergers_2010-1}.
These authors suggested  that the discrepancy could be due to an
overestimation of the $G-M_{20}$ merger rates because of a large
contamination of \protect{non-merging} system or to an underestimation of
$\left<T_\textnormal{obs}\right>$. Alternatively, the theoretical models could
be underestimating the frequency of minor mergers. Our results  suggest that
the $G-M_{20}$ method recovers the intrinsic merger fraction, favouring the
idea that $\left<T_\textnormal{obs}\right>$ are well estimated using the
described methods. Since $G-M_{20}$ morphologies are both sensitive to
minor and major mass ratios. Figure~\ref{fig:correction_fraction} also suggests
that a factor 0.625 correction can be used to determine the rate of exclusively
major mergers from $G-M_{20}$ morphological studies.

\begin{figure} \centering
    \includegraphics[]{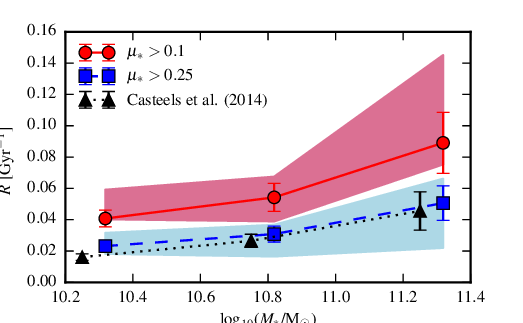}
    \caption{Galaxy merger rate as a function of descendant stellar
      mass estimated from our analysis of  morphological $G-M_{20}$
      for  major mergers ($\mu_* > 0.25$, blue squares) and minor
      mergers ($\mu_* > 0.1$, red circles).  Error
      bars show  Poisson errors.  
The shaded regions represents intrinsic merger rate calculated from
the I-1 simulation for the
corresponding mass ratios.
For
      the sake of comparison,  the observational major merger rates
      derived using a morphological asymmetry method by
      \citet{casteels_galaxy_2014} (black triangles).}
    \label{fig:rate_mass}
\end{figure}

In Figure~\ref{fig:rate_mass} we show the major ($\mu_* > 0.25$) and total
($\mu_* > 0.1$) merger rates of I-1 at $z=0$ as a function of descendant
stellar mass. The blue dots represent the morphologically derived merger rate
computed according to equation \ref{eq:lotz_rate}, but binning the merger
candidates in stellar mass bins. As can be seen, the morphological derivation
matches the intrinsic merger rate very well. This results further corroborates
the estimated observability time-scales for the G-M$_{20}$ criteria; it is also
compatible with results indicating that descendant mass does not affect
G-M$_{20}$  observability time-scales \citep{lotz_major_2011} which implies
that no stellar mass bias is introduced in the merger rate derived using this
method, at least in the $M_* > 10^{10}$ M$_\odot$ range. Blue squares show the
morphologically derived merger rate after applying the correction factor of
0.625 corresponding to major mergers, as can be seen, this correction results
in a good match between the morphologically derived major merger rate as a
function of descendant stellar mass and the intrinsic major rate. Finally, black
triangles correspond to observations of the major merger rate by
\citet{casteels_galaxy_2014} derived using a similar morphological method as
the one shown in the present work, but based on asymmetry and the merger time
scales from \citet{conselice_early_2006}, rather than G-M$_{20}$. The
Illustris simulation is in good agreement with this observations, as also
noted by \citet{rodriguez-gomez_merger_2015}.

\section{Discussion and Conclusions}

We studied non-parametric morphologies of $z=0$ simulated galaxy mergers
in the cosmological context provided by the main Illustris simulation.
From the publicly available idealized mock images, we produced mock g-band
images comparable to the SDSS main galaxy survey. Then we characterized their
morphologies by computing the following non-parametric morphology indicators:
Gini, $M_{20}$, asymmetry, concentration and clumpiness for galaxies with
$M_*> 10^{10}$ M$_\odot$. Our work allows  us to bridge the gap between
isolated merger simulations that explored a limited range of merger conditions
and large scale simulations, which provide galaxy and merger
properties and frequencies in agreement within the current cosmological
paradigm.

We analysed the non-parametric morphologies of galaxies with recent
mergers (within the last 2 Gyr) and with a close companion within 100
$h^{-1}$ kpc. The non-parametric morphologies were correlated with the merger
history and  pair relative velocity and separation, respectively. 

We also analysed the potential sources of contamination that affected a
morphologically selected sample of merging galaxies in the Illustris
simulation. We found that galaxies presenting an artificially
generated ring-like structure constituted a large source of
contamination, followed by lower
mass irregular galaxies, with starbursting appearance. Edge-on galaxies and
bursty spirals were other minor source of contamination. This is also
commonly found in observational samples similarly selected. We were able to
generate a clean sample of morphologically selected galaxies by removing these
sources of contamination.

We found that $G-M_{20}$ morphologies of the complete galaxy sample
reproduced well the trends previously reported by \citet{snyder_galaxy_2015}
for individual subhalos of the Illustris simulation: bulge-dominated galaxies
are located at high $G$, low $M_{20}$ values, while disc-dominated galaxies
are found at low $G$, high $M_{20}$ values.

From the analysis of around {$\sim 40000$} galaxies, we found close
galaxies pairs ($d < 20$ $h^{-1}$ kpc) have
a larger probability to be selected by $G-M_{20}$
morphologies ($\sim 20$ per cent) and they are also well-selected by the
asymmetry $A$. The analysis of the completeness shows that 50 per cent
and 35 per cent of galaxies with positive merginess are in pairs with
$d < 45$~ $h^{-1}$ kpc and $\mu_* > 0.25$ and  $\mu_* > 0.1$, respectively.
Major merger remnants constituted the second subsample that the $G-M_{20}$
criteria was able to better differentiate, $\sim 5$ per cent of major merger
remnants show perturbed morphologies.
Nevertheless, $\sim 98$ per cent
of the galaxies above the demarcation line have experienced a
perturbation (i.e. a close interaction or a recent merger). 

In agreement with previous works, the largest fraction of merger remnants
and galaxy pairs are located below the demarcation line. However, the
merginess is found to capture signatures of their actual state of disturbance.
A clear correlation between the merginess and the elapsing time to the latest
merger event is found. This trend is stronger for major merger events, and
gets weaker when merger remnants of smaller mass ratios are included. However,
all of them show merginess larger than the average of the whole sample.

Using the observability time-scales from isolated merger simulations by
\citet{lotz_galaxy_2008,lotz_effect_2010,lotz_effect_2010} and the mass ratios
and gas fractions distribution of $z=0$ mergers in the Illustris main
simulation, we computed the average observability time-scale of the
simulation. We found a value of $\left<T_\textnormal{obs}\right> \sim 0.2$
Gyr, very similar to other reported values from simulations and semi-
analytical models \citep{lotz_major_2011}. Next, we put this value to the
test, comparing the intrinsic merger rate of Illustris to the merger rate that
would have been derived from morphological studies using the $G-M_{20}$
criterium. We found that after accounting for the contamination of
morphologically selected galaxies, no further corrections where necessary to
reconcile the intrinsic merger rate to the morphologically derived one.  This
agreement indicates that the computed average observability time-scales are a
correct estimation of the time that $z=0$ mergers are detected above the
demarcation line. This result  validates the findings obtained by the isolated
merger simulations and shows that the cosmological context of galaxy formation
does not introduce effects that greatly alter the observability time-scales of
merger events at least in the local Universe.

We notice that the discrepancy found by \citet{lotz_major_2011} between the
observed total merger rate (minor plus major mergers) and the intrinsic merger
rate from simulations \citep{hopkins_mergers_2010,somerville_semi-analytic_2008,stewart_galaxy_2009} is maintained in the Illustris simulation.
Observational estimations indicate a large merger rate (0.37 Gyr$^{-1}$ at
$z=0.3$), while Illustris predicts a much lower rate (0.06 Gyr$^{-1} $at
$z=0$). This large difference can not be explained by reasonable evolutionary
trends of the merger rate with redshift. \citet{lotz_major_2011} proposed
several possible solutions to this discrepancy: simulations could be
under-predicting the merger rate of minor mergers, the observability time-
scales of $G-M_{20}$ morphologies could be underestimated or observations
could be affected by large contamination from non-merging systems. We have
shown that computing average observability time-scales using the standard
procedure results in correctly derived observed merger rates,  compatible to
the intrinsic merger rate. We also note that, completely disregarding
contamination sources in our estimations of the morphological derived rates
produces  a merger rate of 0.2 Gyr$^{-1}$, closer to the reported
observational values. Our results suggest that contamination sources might
explain this discrepancy, which seems to be more likely associated to minor
merger events. Indeed, major merger rates are well reproduce by the I-1
simulation when compared to non-parametric morphological studies using the
asymmetry statistic \citep{casteels_galaxy_2014} and we have shown that
$G-M_{20}$ morphologies can reproduce the major merger rate if a correction
factor of $\sim 0.63$ is applied.

We also studied the effects of mergers on the asymmetry statistics $A$. We
found that asymmetry increased for close pair with  $d < 35$ $h^{-1}$ kpc .
However we found that asymmetry greatly increases for lower mass galaxies,
resulting in most galaxies with $M_* < 10^{10.5}$ M$_\odot$ with $A > 0.35$,
regardless of the presence or absence of any merging event. Based on previous
works, we suggest that this  effect might be  largely caused by the
combination of low numerical resolution for low mass galaxies and stochastic
formation of stellar particles that greatly affect the appearance of galaxies.
Although  this renders the asymmetry statistic of the simulation very hard to
properly compare with observations, as this effect is intrinsic to
simulations, we found the trend with $A$ is present at all mass intervals. The
larger changes of $A$ is detected for higher stellar mass galaxies in pairs
with $d < 30$ $h^{-1}$ kpc . Smaller galaxies show an increasing level of $A$
at all relative distances which remains to be confirmed with higher numerical
resolution simulations. We also analysed the dependence of $A$ with relative
velocities, galaxies with  $V < \sim 300$ km s$^{-1}$ have $A > 0.35$,
while pairs with higher relative velocities have $A < 0.35$, suggesting that
flyby events have no significant impact on morphological disturbances.

\section*{Acknowledgements}

The authors acknowledge the grants PICT 2011-0959 from Argentinian ANPCyT, and
PIP 2012-0396 from Argentinian CONICET and Fondecyt Regular 115033, Southern
Astrophysics Network Redes Conicyt 150078 and proyecto interno MUN UNAB 2015.
This research made use of Astropy
\citep{astropy_collaboration_astropy:_2013}, numpy \citep{walt_numpy_2011},
and matplotlib \citep{hunter_matplotlib:_2007}

\bibliographystyle{mnras}
\bibliography{draft_morphology}

\appendix
\section{Numerical approach: comparison to DIGGSS morphologies}
\label{sec:appendix}

{\em ''Dusty Interacting Galaxy GADGET-SUNRISE Simulations''}
(DIGGSS)\footnote{\url{http://archive.stsci.edu/prepds/diggss}} are series of
isolated merger simulations used to derived the non-parametric merger \protect
{time-scales} in \citet{lotz_galaxy_2008,lotz_effect_2010,lotz_effect_2010-1}.
In order to validate and calibrate our approach to the computation of \protect
{non-parametric} morphologies, we recalculated those of the DIGGSS mock images
and compared them to the tabulated results.

\begin{figure*}
    \centering
    \includegraphics[width=0.8\textwidth]{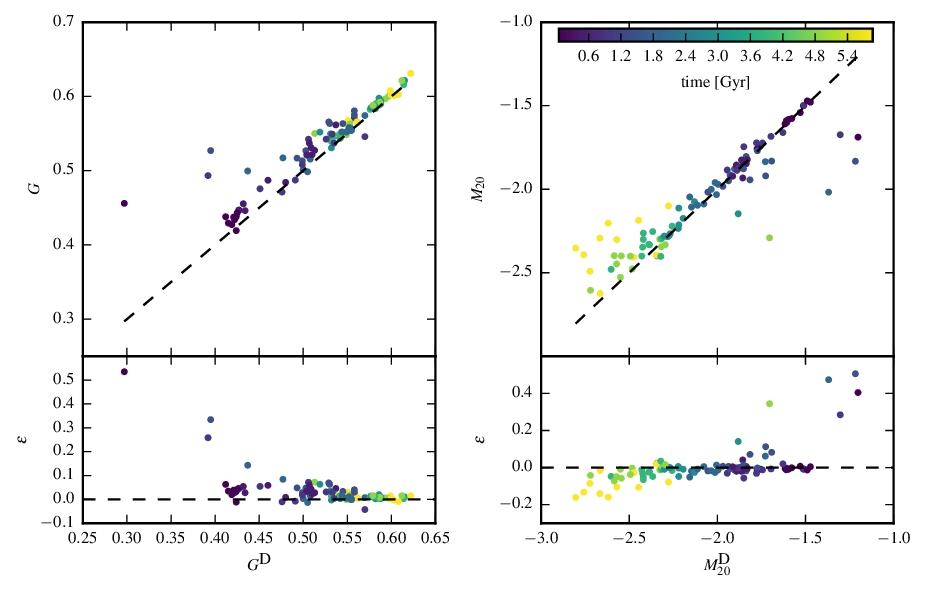}
    \caption{Left: gini computed using the methods described in this paper vs. tabulated values for the DIGGSS g3iso galaxy. Right: the same for the $M_{20}$ statistic. Colours represent the time since the start of the simulation. In both cases, lower panels represent the relative deviation of the parameters computed using our method to the tabulated values.}
    \label{fig:lotz_comparison_gini}
\end{figure*}

Figure~\ref{fig:lotz_comparison_gini} compares our results for the Gini and
$M_{20}$ statistics to the tabulated values for the g3iso galaxy, whose
characteristic are listed in table \ref{tab:g3iso}. Each point in the figure
represents a mock image of the galaxy taken at a certain time in the
simulation, with one of the eleven cameras distributed around the galaxy. The
lower panels  represent the relative deviation from the tabulated value. As
can be seen a good general agreement was found, most of our estimations lie
within 10 per cent of the tabulated value, with a few outliers for the Gini
statistic at low $G$, and  for $M_{20}$ at low and high values.

\begin{table}
    \centering
    \caption{DIGGSS g3iso fundamental parameters.}
    \begin{tabular}{ll}
        \hline
        Virial mass & $1.2 \times 10^{12}$ M$_\odot$ \\
        Dark matter halo concentration & 6 \\
        Baryonic mass & $6.2 \times 10^{10}$ M$_\odot$ \\
        Mass of stellar disc & $4.1 \times 10^{10}$ M$_\odot$ \\
        Mass of stellar bulge & $8.9 \times 10^{10}$ M$_\odot$ \\
        Mass of gaseous disc & $1.2 \times 10^{10}$ M$_\odot$ \\
        Fraction of baryons in the bulge & 0.14 \\
        Fraction of baryons in gas & 0.19 \\
        Scalelength of stellar disc & 2.85 kpc \\
        Scalelength of bulge & 0.62 kpc \\
        Scalelength of gaseous disc & 8.55 kpc \\
        \hline
    \end{tabular}
    \label{tab:g3iso}
\end{table}

\begin{figure*}
    \centering
    \includegraphics[width=0.8\textwidth]{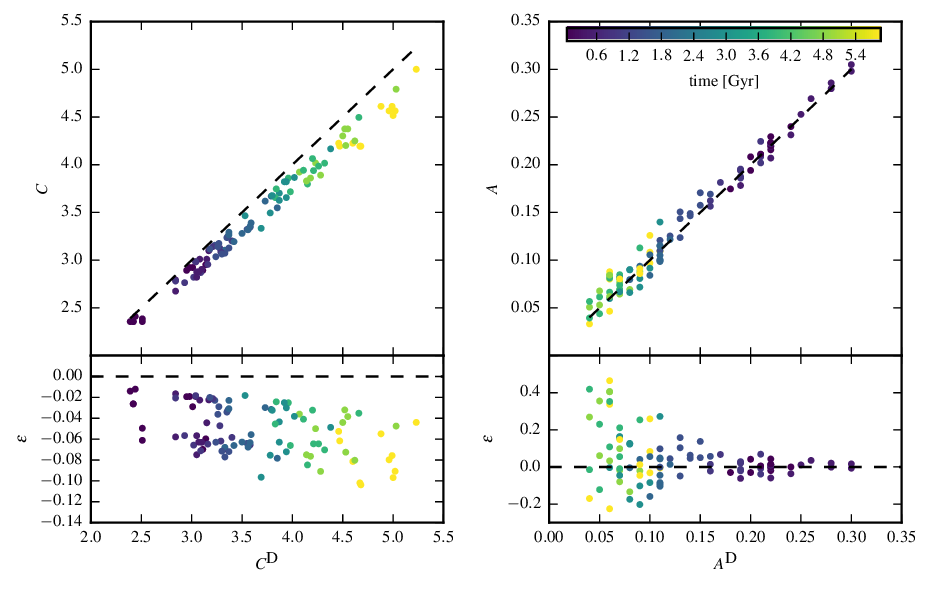}
    \caption{Left: concentration computed using the methods described in this paper vs. tabulated values for the DIGGSS g3iso galaxy. Right: the same for the asymmetry statistic. Colours represent the time since the start of the simulation. In both cases, lower panels represent the relative deviation of the parameters computed using our method to the tabulated values}
    \label{fig:lotz_comparison_a}
\end{figure*}

Similarly, Fig~\ref{fig:lotz_comparison_a} compares our results for the
concentration and asymmetry. The Concentration presents a slight bias towards
lower concentration values that increases towards higher concentrations, but
all results are within 10 per cent of the tabulated values. The asymmetry
agrees well with the tabulated values specially at higher asymmetry values.
For low asymmetry there is larger dispersion.

These results show that our approach is able to produce robust non-parametric
morphologies comparable to the ones derived in previous works.

\bsp    
\label{lastpage}
\end{document}